# Effects of Al Addition on the Native Defects in Hafnia

Quan Li, K. M. Koo, W. M. Lau
Department of Physics, The Chinese University of Hong Kong, Shatin, New Territory, Hong Kong

P. F. Lee, J. Y. Dai
Department of Applied Physics and Materials Research Center, The Hong Kong Polytechnic University, Hung Hom, Kowloon, Hong Kong

Z. F. Hou, X. G. Gong
Surface Physics Laboratory & Department of Physics, Fudan University, 200433-Shanghai, China

**Abstract**

Two occupied native defect bands are detected in pure $HfO_2$, with one located in the middle of the band gap and the other slightly above the valence band maximum. The investigation on the electronic structures of hafnium aluminate thin films as a function of Al concentration discloses the evolution of such bands—while the density of states of the former one reduces drastically with the Al addition, that of the later one remains rather unaffected. Our first principles studies of the system attribute the two bands to the charged oxygen vacancy, and the oxygen interstitial related defect states of the $HfO_2$, respectively. We further demonstrate that the observed evolution of the defect bands originates from the interaction in-between the added Al and the native defects of pure $HfO_2$, which effectively passivates the $V_O^+$ induced mid-gap states but has little effect on other aspects of the electronic structure of the material.

*Corresponding authors:   Dr. Quan Li: liquan@phy.cuhk.edu.hk
                          Dr. X. G. Gong: xggong@fudan.edu.cn

The scaling requirements in the complementary metal-oxide-semiconductor (CMOS) integrated circuit technology predict the inadequacy of $SiO_2$ as gate dielectric material for system smaller than 0.1 μm, which leads to an extensive search of the high-dielectric constant (high-K) materials as potential replacements for $SiO_2$ [1]. Among these high-K dielectrics, pseudo-binary Hf-based materials appear to be one of the most promising material systems and they have indeed demonstrated various desirable electrical behaviors. In such a pseudo-binary system, the addition of a third element (usually Si or Al) to $HfO_2$ is expected to increase, relative to the properties of pure $HfO_2$, the Si/dielectric interfacial stability [2,3], crystallization temperature [4], band gap, and both the valance and conduction band offset to Si [5-7].

On the other hand, there are still practical deficiencies in the $HfO_2$ system. In particular, various native defects in $HfO_2$, such as oxygen vacancy and interstitial, have been both theoretically predicted [8-11] and experimentally observed [12, 13]. Once formed, these defects not only introduce defect levels in the band gap, but also serve as charge traps. They can thus severely affect the electrical behavior of the materials, such as leakage current and charge scattering [8,14,15]. These effects become even more complicated in the pseudo-binary system because the aforementioned addition of the third element in $HfO_2$ may interact with the native defects of $HfO_2$, modify the defect states, and thereby either improve or deteriorate the electrical performance of the gate dielectrics. To date, little is known about the interaction between the additional element and the native defects in $HfO_2$, despite its significance in understanding and predicting the electrical behavior of the pseudo-binary Hf-based gate dielectrics.

To address these outstanding issues, we have carried out a systematic investigation on the electronic structure changes of a representative pseudo-binary Hf-based material—hafnium aluminate as a function of aluminum concentration. In this letter, we report our experimental detection of the presence of at least two broad occupied defect bands in the bandgap of the nominally pure $HfO_2$ films (referred as pure $HfO_2$ therein), experimental measurements of their individual changes in density of states as a function of aluminum addition, and our theoretical explanation of these observations with ab initio studies of the relevant compounds. The effects of the defect properties on the electrical behavior of the materials are also discussed. The current work sheds light on the general understanding of the defect electronic structure of HfO$_2$, and provides a novel approach that could eliminate some of the native defect states and improve the electrical performance of the gate dielectrics.

The hafnium aluminate thin films with thickness of about 20 nm were deposited by pulsed laser deposition on p-type (100) Si substrates using a high purity hafnium aluminate target with different Hf/Al ratio (Table I). The silicon substrates were treated by an HF-etch just before thin film deposition, an etch which is known to leave silicon surface terminated by hydrogen. The XPS measurements were carried out using a PHI Quantum 2000 system, equipped with a monochromatized Al $K_\alpha$ source, a surface charge neutralizer, and a low-energy argon ion gun. All high-resolution scans were taken at a photoelectron take-off angle of 90° and calibrated to Au 4f$_{7/2}$ at 84.0 eV from a sputter-cleaned gold foil. Measurements of the Fermi level position in the bandgap of were also calibrated by the threshold of photoemission of the sputter-cleaned gold reference under the same spectral resolution conditions. The microstructures of the thin films were investigated by transmission electron microscopy (TEM, Tecnai 20ST). The valence electron energy loss measurements of the films were performed using the Gatan GIF system attached to the TEM with an energy resolution of 0.7 eV. The spectra were acquired in the diffraction mode at small momentum transfer with an angular resolution about 0.2 mrad.

The Hf/Al ratios of the films are determined from the XPS survey scan using the corresponding peak areas, as illustrated in Table I. Scanning transmission electron microscopy and transmission electron diffraction suggest similar thickness (~20 nm) and morphology of the all as-deposited films—the films are amorphous and no interfacial layer is detected.

The loss functions ($\text{Im}[-\frac{1}{\varepsilon}]$) of the films (Figure 1) are obtained by removing the plural scattering from the corresponding energy loss spectra using direct deconvolution method [16]. The band gap of pure HfO$_2$ is estimated to be ~6.0 eV, which is consistent with the literature reports. More accurate determination of the band gap is difficult, mainly due to the existence of the mid-gap state(s). Little change of the band gap is identified in the as-deposited films as the Al concentration increases, until pure alumina is obtained, for which the band gap is estimated as ~6.5 eV, also consistent with the literature value of amorphous alumina [7]. An obvious mid-gap band with excitation energy of 3 to 6 eV appears in the pure $HfO_2$ films. Such a feature significantly decreases when aluminum is introduced in the films, and completely diminishes in the pure alumina film.

Detailed information of the occupied mid-gap states above the valence band maximum (VBM) are obtained using high-resolution XPS scan in the low energy range (0-10 eV), which is known to provide a much better energy resolution and signal to noise ratio in the low energy range, compared to the EELS measurements. The most predominant features in the spectra are the two mid-gap bands (centered at ~2.5 eV (Band I) and ~0.5 eV (Band II) above the VBM) observed in pure $HfO_2$ (as marked by arrows in Figure 2). It is interesting to note that the two bands behave differently with the Al addition. The intensity of Band I decreases drastically as the Al concentration increases in the films, and becomes un-detectable when the Hf/Al ratio reaches 1/6. In comparison, little change is observed for the intensity of Band II as the film composition varies, and such band only disappears in pure alumina.

The complementary information provided by the transmission EELS and the XPS measurements give a complete picture of the defect electronic structures in pure $HfO_2$ and their evolution with the Al addition. Nevertheless, we should note that loss of detailed information is expected in the EELS due to its poor energy resolution, and defect states at low concentration may not be observed due to the detection limits of both experimental methods.

The observed bands centered at ~2.5 eV and ~0.5 eV above the VBM must result from the defect states in pure $HfO_2$. The most commonly reported native defect states in $HfO_2$ are related to neutral/charged oxygen vacancy and interstitial, which contribute to the electrical behavior of the gate dielectric material [8,14,15]. Intuitively, the change of the defect bands observed in both EELS and XPS must be related to the Al addition. In order to understand the origin of such change and predict possible electrical behavior of the films after the Al addition, we investigate the interaction between the Al and the original defect states and the effect of such interaction on the electronic structure of the film, using density functional theory (DFT).

Our calculations were based on DFT with the generalized gradient approximation (GGA) [17] as implemented in the VASP code [18], in which the valence electrons are treated explicitly and their interactions with the ionic core are described by the ultra-soft pseudopotentials [19]. The energy cutoff is 500 eV. To simulate the defect structures, we used a 2 x 2 x 2 supercell with 96 host atoms, and only Γ point is employed in the calculation of the structural relaxation.. Atoms are removed from or added to the supercell in order to model the vacancy or interstitial defects, respectively. For the simulation of the Al subsitutional defect ($Al_{Hf}$), one of the Hf atoms in the supercell is substituted by an Al atom. The defect pairs ($V_O$-$Al_{Hf}$, $O_i$-$Al_{Hf}$, or $O_2$-$Al_{Hf}$) are formed by a substitutional Al atom and an adjacent oxygen vacancy or interstitial. In the calculations, the lattice vectors of the supercell were fixed at equilibrium value for pure $HfO_2$ and all atoms are allowed to relax until atomic forces were less than 0.01 eV/ Å. In the calculation with charged defects, we apply a uniform background charge to neutralize the supercell [20]. We identify the defect levels in $HfO_2$ by calculating the ionization energies and electron affinities with respect to the bottom of the conduction band of $HfO_2$. The vertical ionization energies (VIP) ($I_p(D^q)$) and relaxed electron affinities(REA) ($\chi_e(D^q)$) of the defects are computed by following the Eq. (2) and (3) in Ref. [8]. Our calculated band gap $E_g^{theor}$ = 3.86 eV using the procedure of Eq.(1) in Ref.[ 22] is smaller than the experimental value of band gap $E_g^{exp}$ = 5.68 eV [21], The defect ionization energies $I_p(D^q)$ and electron affinities $\chi_e(D^q)$ are then corrected using $\kappa$ = $E_g^{exp}$ - $E_g^{theor}$ = 1.82 eV and procedure described in Ref. [8].

The calculated ionization energies and relaxed electron affinities of these defects are summarized in Table II. Those values of the pure $HfO_2$ are in good agreements with the literature reports [8]. The formation of a neutral oxygen vacancy introduces a fully occupied defect level with two electrons ($V_O^0$) in the middle of the band gap. The highest occupied state is located at 2.02 eV above the VBM as indicated by our calculated ionization potential (Table II). The lowest unoccupied state of the relaxed neutral vacancy is located at 1.84 eV below the conduction band minimum (CBM). If oxygen vacancies are the only defects present in $HfO_2$, the Fermi energy are expected to be in between of these donor transition energy levels and CBM at finite temperature. Furthermore, when defect concentration is high, defect bands can form centered around these transition energy levels.

The addition of neutral oxygen interstitials ($O_i^0$ and $O_2^0$) has two effects in general. First, the defects contribute to occupied and unoccupied defect states in the bandgap. Second, they may also interact with $V_O$ and thereby alter the physical properties of oxygen vacancy. Table II and Fig. 3 show that the VIP of $O_i^0$ and $O_2^0$ are 0.04 and 0.08eV smaller than the band gap. Their respective REA are at 3.71 and 4.64 eV. Since the unoccupied states are below the highest occupied states of both $V_O$ and $V_O^+$, they can draw electrons from the oxygen vacancies and form $O_i^-$ and $O_2^-$. In fact, the data in Table II also show that the reactions of $V_O + O_i^0 (O_2^0) \rightarrow V_O^+ + O_i^- (O_2^-)$ are indeed energetically favorable, formation of $O_i$ ($O_2$) will compensate $V_O$, removing electrons from its defect level. The finally resulting $O_i^{2-} (O_2^{2-})$ states have VIP that are ~0.66eV smaller than the VBM, whereas the $V_O^{2+}$ has a REA at about 3.26 eV. The Fermi level of the system will drop accordingly. The actual behavior and electrical properties will obviously depend on the actual amounts of these defects in a real sample.

To understand the correlation between the theoretical calculations and the experimentally observed energy states in the band gap, one should note that XPS measures the electron excitations from the occupied states to vacuum and EELS measures the transitions to unoccupied states. To illustrate such correlation, the calculated defect ionization energies are plotted in the schematic energy diagram shown in Figure 3. By comparing the energy levels obtained in theoretical calculation, we propose that the two defect bands observed ~2.5 eV and ~0.5 eV above the VBM in the XPS spectrum of the pure $HfO_2$ films can be assigned to the oxygen vacancy and interstitial related defects, respectively. These results are consistent with the relevant theoretical and experimental data in the literature [8, 12]. From our experimental XPS data, we know that the actual Fermi level position in our pure HfO$_2$ film is at about 4 eV above VBM. This implies that the amount of V$_O$ must be greater than that of oxygen interstitials (O$_i^0$ and O$_2^0$). The aforementioned defect interactions among the vacancies and interstitials would therefore leave V$_O^+$ as a half occupied defect states with the smallest VIP in the band gap of pure HfO$_2$, and thus a corresponding Fermi level in-between the energy level of V$_O^+$ and the CBM.

Regarding the effects of Al addition, because Al has three valence electrons and Hf has four, substitution of Al at Hf site will create a hole states near the VBM. This acceptor level will compensate the V$_O$ donor state, remove electrons from the donor level. When Al concentration is more than twice as large as the oxygen vacancy concentration, all the electron in the oxygen vacancy donor levels will be removed, i.e., V$_O$ is 2+ charged. Indeed this change is clearly demonstrated by the observed diminishing Band I with Al addition in the XPS data in Figure 2. Although the Al addition also changes other occupied defect states near the VBM, the changes on their VIP are relatively small. Furthermore, Al$_{Hf}$ itself also creates occupied defect levels near VBM contributing to the XPS signal. This explains the persistent appearance of Band II in the XPS spectra of the HfO$_2$ films with Al addition (Figure 2). In short, our experimental and theoretical results suggest that ionized oxygen vacancies and interstitials are the major defects observed in our experiments. This is also supported by the C-V measurement of the corresponding films, which suggest the existence of large amount of positively charged defects [14].

Considering the excitation from the fully-occupied/half-occupied states, the V$_O^+$ state is located in the middle of the band gap. Large concentration of such state would significantly reduce the effective band gap and high-$K$/Si valence band (VB) offset. This makes the material inappropriate to work as gate dielectric. The addition of Al would eliminate the half occupied states in the band gap, which is an effective solution to the problem. As a comparison, the charged O interstitial defect states are located close to the valence band maximum (VBM). Large amount of such defects in the films would result in a slight reduction of the band gap and VB offset from that of the Si. The addition of Al has little effect on the energy levels of such defects. Due to the relatively large band gap and VB offset between $HfO_2$ and Si, this would not lead to much deterioration of the electrical behavior of the gate dielectric.

On the other hand, the unoccupied/half-occupied states of these defects could serve as electron traps, which not only serve as charge scattering centers but also affect the leakage current of the dielectric film. $V_O^+$ is a half-filled state, its electron affinities is relatively large, indicating that it can serve as effective electron traps. When interacting with Al in the neighborhood, the electron affinity of the complex is reduced (Table II).

In conclusion, being one of the most promising candidates to replace $SiO_2$ and serve as the next generation of gate dielectric material in the CMOS industry, the properties of $HfO_2$ are degraded by the presence of various native defects because they modify the material's electronic structures and deteriorate the film's electrical behavior. Understanding the nature of these defects and seeking effective methods to reduce/eliminate these defects is of primary concern. In the presented work, we found that the two defect bands centered at ~2.5 eV and ~0.5 eV above the VBM in pure hafnia can be identified as originating from the oxygen vacancies and oxygen interstitials, respectively. Introducing Al to the pure $HfO_2$ leads to interaction between Al and these defects, which passivates the occupied $V_O$ gap states, but has little effect on other aspect of the electronic structure of the material. The current study, therefore, suggests an alternative approach to eliminate these defects, i.e., through defect engineering—by introducing an appropriate dopant to achieve the goal of gap state reduction and even elimination.

**Acknowledgements**: We thank for the critical comments of DS Wang. XGG is partially supported by the NSF of China, the special funds for major state basic research. The computation was performed at Shanghai Supercomputer Center and Supercomputer Center of Fudan..

**Tables**

Table I. Hafnium/Aluminum ratio in the hafnium aluminate target and the as-deposited thin films

| Sample No. | Hf/Al ratios in the targets | Hf/Al ratios in the as-deposited thin films |
|---|---|---|
| HfAlO1 | 1/0 | 1/0 |
| HfAlO2 | 2/1 | 4/1 |
| HfAlO3 | 1/1 | 1.5/1 |
| HfAlO4 | 1/2 | 1/2 |
| HfAlO5 | 1/4 | 1/6 |
| HfAlO6 | 0/1 | 0/1 |

Table II.–Ionization energies and electron affinities of defects in different charged states.

| Defects | $I_p(D^q)$ | | $\chi_e(D^q)$ | |
|---|---|---|---|---|
| | This work | Previous[a] | This work | Previous[a] |
| $V_O^0$ | 3.66 | 3.41 | 1.84 | - |
| $V_O^+$ | 3.48 | 3.75 | 3.01 | 2.76 |
| $O_i^0$ | 5.64 | 5.55 | 3.71 | 3.95 |
| $O_i^-$ | 5.37 | 5.38 | 5.02 | 4.75 |
| $O_2^0$ | 5.60 | 5.53 | 4.64 | 4.67 |
| $O_2^-$ | 5.42 | 5.46 | 5.13 | 5.06 |
| $(V_O\text{-}Al_{Hf})^0$ | 2.97 | - | 2.61 | - |
| $(V_O\text{-}Al_{Hf})^+$ | 5.75 | - | 2.33 | - |
| $(O_i\text{-}Al_{Hf})^0$ | 5.62 | - | 5.43 | - |
| $(O_i\text{-}Al_{Hf})^-$ | 5.61 | - | 1.76 | - |
| $(O_2\text{-}Al_{Hf})^0$ | 5.47 | - | 5.07 | - |
| $(O_2\text{-}Al_{Hf})^-$ | 5.40 | - | 4.50 | - |

**Figure Captions**

Figure 1. Loss functions of the Hf-Al-O thin films with different Al concentrations showing their band gaps.

Figure 2. XPS low loss spectra of the Hf-Al-O thin films with different Al concentrations.

Figure 3. Single-particle defect energy level diagram showing the effect of Al addition on the ionization energies for various defects related to the threefold coordinated oxygen in monoclinic hafnia. All energies are in eV. The occupation of the energy levels is given by the number of small ↑(spin up) and ↓(spin down) arrows, respectively.

Figure 1

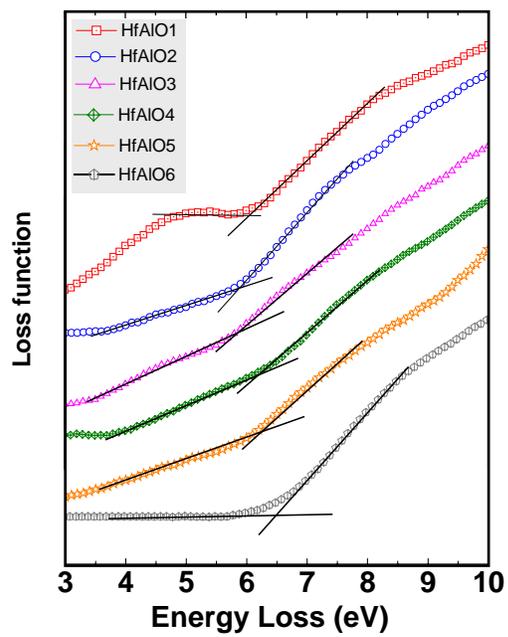

Figure 2

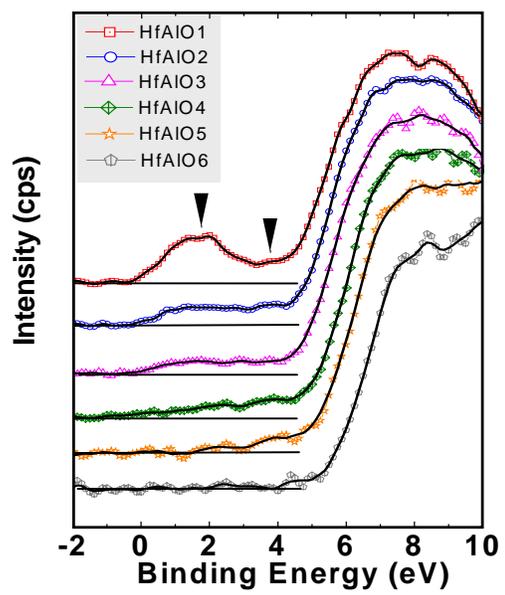

Figure 3

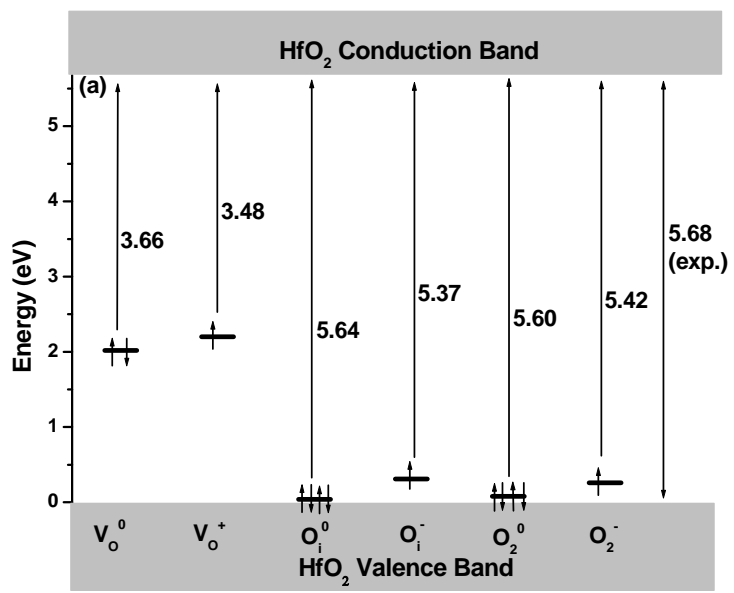

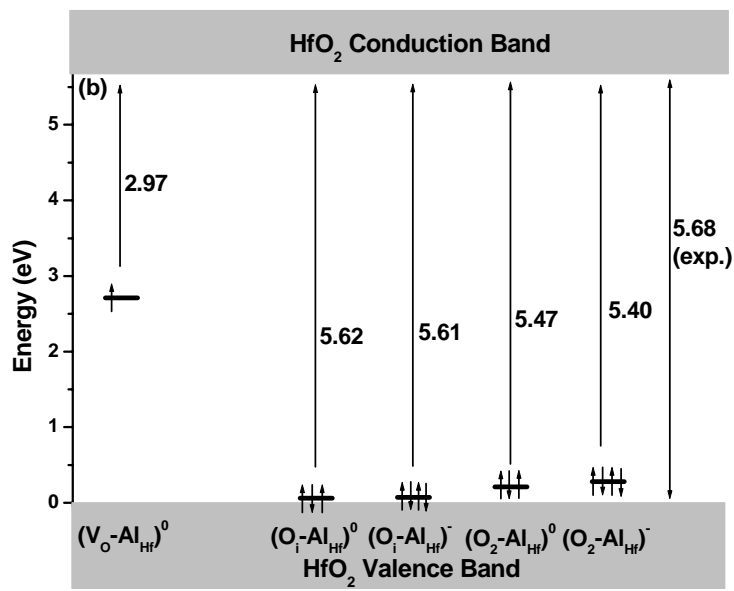